\documentclass[12pt]{article}
\usepackage[top = 2 cm, bottom = 3 cm, left = 2.5 cm, right = 1.8 cm]{geometry}
\usepackage{authblk, cite, color, amssymb, amsmath}
\usepackage[colorlinks = true, linkcolor = blue, citecolor = red]{hyperref}
\usepackage[dvipsnames]{xcolor}

\begin{document}

\title{Electroweak Phase Transitions in Einstein's Static Universe}

\author{\bf M. Gogberashvili}
\affil{\small Javakhishvili Tbilisi State University, 3 Chavchavadze Avenue, Tbilisi 0179, Georgia \authorcr
Andronikashvili Institute of Physics, 6 Tamarashvili Street, Tbilisi 0177, Georgia}

\maketitle

\begin{abstract}
We suggest using Einstein's static universe metric for the metastable state after reheating, instead of the Friedman-Robertson-Walker spacetime. In this case strong static gravitational potential leads to the effective reduction of the Higgs vacuum expectation value, which is found to be compatible with the Standard Model first order electroweak phase transition conditions. Gravity could also increase the CP-violating effects for particles that cross the new phase bubble walls and thus is able to lead to the successful electroweak baryogenesis scenario.

\vskip 1mm

\end{abstract}
\vskip 5mm


According to standard cosmology at the electroweak scale our universe is in radiation dominated phase where all standard model particles are massless \cite{Much}. Once the temperature drops below the critical value, $T \sim 170$~GeV, the electroweak phase transition has occurred and the Higgs boson, gauge bosons and fermions (except of neutrinos) acquire masses through the Higgs mechanism. The order of this phase transition depends on the details of the Higgs potential with the temperature dependent terms \cite{DLHLL-1, DLHLL-2, Moss, Katz-Per}. To have a first order phase transition effective Higgs potential of the model should have several minima. Of special importance is the cubic term in effective potential, which is essential to generate a potential barrier between the symmetric and broken phases and thus can provide that the phase transition to be of the first order. In the Standard Model the cubic term, $ET^3$,  is contributed only by the electroweak gauge bosons. If at zero temperature the Higgs field at the minimum of the potential has the value
\begin{equation} \label{v}
v \approx 246~{\rm GeV}~,
\end{equation}
the parameter $E$ is the cubic term of the effective potential is of order of
\begin{equation} \label{E}
E \approx \frac{2M_W^3 + M_Z^3}{4\pi v^3} \approx 0.01 ~,
\end{equation}
where $M_W$ and $M_Z$ the gauge bosons masses. Then the condition that the Higgs effective potential has two minima leads to the very small value for the Higgs self-coupling parameter,
\begin{equation}\label{l=2E}
\lambda \approx 2E \approx 0.02~,
\end{equation}
which according to (\ref{v}) is incompatible with the observed Higgs boson mass,
\begin{equation}
m_H = \sqrt{2\lambda} v \simeq 125~GeV ~.
\end{equation}
This means that within the minimal Standard Model the electroweak phase transition is a smooth crossover, or of the second order \cite{Moss, Katz-Per}.

From the other hand, the first-order electroweak phase transitions may solve some cosmological problems, like the generation of the baryon asymmetry of the universe (see recent reviews \cite{Mor-Ram, Kons}). A first order cosmological phase transition proceeds through the formation and expansion of cosmic bubbles. In this scenario spacetime is separated into two manifolds with their own distinct metrics, which are typically joined across a thin wall (domain wall). The dynamics of such objects can be very complicated, depending on the matter content of the interior (true vacuum) and exterior (false vacuum) regions as well as the tension on the bubble walls and how they interact with the surrounding plasma \cite{BKT87}. In bubble models matter-antimatter asymmetry can be generated at the electroweak scale, because all three Sakharov conditions (baryon number violation, $C$ and $CP$ violation and departure from the thermal equilibrium) are fulfilled. Baryon production is a bubble surface effect but can be used to explain the observed matter-antimatter asymmetry, since the entire universe is swept out during the first order phase transition and we may live in one bubble today \cite{CKN}. However, in traditional approach not only the electroweak phase transition is not of first order, but the $CP$-violation from the Cabibbo-Kobayashi-Maskawa matrix is too small, and to obtain the observed baryon asymmetry various extensions of the Standard Model have been proposed \cite{Moss, Katz-Per, Mor-Ram, Kons}.

We want to emphasize that the conventional scenario of electroweak baryogenesis do not take into account gravitational effects. It is assumed that at the electroweak scale the universe was radiation dominated (filled with relativistic particles in thermal equilibrium) and the Friedmann-Robertson-Walker scale factor was unimportant for the particle reactions \cite{Much}. However, details of the anticipated periods (inflation and reheating) and conditions at the moment of transition to the radiation dominated phase, with the normal expansion of spacetime, are still poorly understood. For instance, if one assumes the existence of black holes in the early universe, the gravitational effects in their static fields are able to modify parameters of the effective Higgs potential and lead to the first order phase transitions \cite{BGM, Tetra}.

It is known that the Cosmological Principle for the uniform matter distribution leads not to the Friedmann-Robertson-Walker non-static solution only, but gives the Einstein's static universe as well \cite{Haw-Ell}. Einstein's static universe refers to the homogenous and isotropic universe with the positive cosmological constant and positive spatial curvature. In the framework of General Relativity this model has been widely investigated for several kinds of matter sources (see \cite{Bar-Tsa, Bar-Yam} and references therein). The metric of Einstein's universe can be written in the form:
\begin{equation} \label{EU-metric}
ds^2 = dt^2 - \frac {dr^2}{1 - \Phi (r)} - r^2 \left( d\theta^2 + \sin^2 \theta d\phi^2 \right)  ~,
\end{equation}
where
\begin{equation} \label{Phi}
\Phi (r) = \frac {8\pi}{3} G \rho r^2~,
\end{equation}
denotes the gravitational potential and $\rho$ is the uniform cosmic fluid density. In (\ref{EU-metric}) it has been set $g_{tt} = 1$, since in cosmological case there must be a universal proper time for all fundamental observers. Introducing the radial coordinate transformation,
\begin{equation}
r = \frac {R}{1+ 2\pi G\rho R^2/3}~,
\end{equation}
the metric (\ref{EU-metric}) can be written also in terms of Cartesian coordinates \cite{Tolman}:
\begin{equation} \label{EU-x,y,z}
ds^2  = dt^2 - \frac {1}{\left(1 + 2\pi G\rho R^2/3\right)^2} \left( dx^2 + dy^2 + dz^2 \right)  ~.
\end{equation}
It was found that Einstein's static universe (\ref{EU-metric}) is critically unstable to gravitational collapse or expansion and in modern cosmological models usually is considered only as the initial state for the inflationary phase \cite{Ell-Maa}.

In our opinion the short metastable period with $\ddot a = \dot a = 0$, in course of transition of the inflation ($\ddot a > 0$) into the deceleration stage ($\ddot a < 0$), i.e. during or after the reheating epoch, when our universe was a spherule with the homogenous cosmic fluid of all kinds of ultrarelativistic particles, is natural to describe by Einstein's static solution (\ref{EU-metric}), with the later transition to the non-stationary radiation dominated phase with the Friedmann-Robertson-Walker metric. This can be achieved by introducing a dark energy like substance, presence of which at early times is a quite generic feature of several dynamical dark energy models \cite{Wett}, e.g. $f(R)$ modification of gravity \cite{f(r)-1, f(r)-2}, or the decaying scalar field (quintessence) \cite{CST}, rather than a large fixed value classical cosmological constant.  One can also assume that the higher-energy degrees of freedom of the quantum vacuum during reheating do not cancel the contribution of the zero-point motion of the quantum fields and the nullification of vacuum energy in the equilibrium vacuum is not acquired at this stage.

Introduction of Einstein's static metric (\ref{EU-metric}) for our spherule-universe can radically change the common view that for the observed large Higgs mass the cosmological electroweak phase transition is of the second order \cite{Moss, Katz-Per}. In a static island of space we can expect appearance of large gravitational potential in the Standard Model Lagrangians, similar to the case with a static black hole background  \cite{BGM, Tetra}. Indeed, introducing the density function, $\Omega = \rho /\rho_c$, where
\begin{equation}
\rho_c = \frac {3}{8\pi G d^2}
\end{equation}
is the critical density parameter and $d$ is the horizon distance at the electroweak scale, the gravitational potential (\ref{Phi}) can be written in the form:
\begin{equation} \label{Phi-R}
\Phi (r) = \Omega ~\frac {r^2}{d^2}~.
\end{equation}
A preferred location in the universe is absent and for uniform matter distribution we expect the existence of a constant average gravitational potential, $\langle \Phi (r) \rangle$, and the radial dependence will disappear. The total matter density of our universe at all stages of its evolution is assumed to be close to unity \cite{Planck},
\begin{equation}
\Omega \lesssim 1.005~.
\end{equation}
Then the average gravitational potential (\ref{Phi-R}) in the Einstein static universe (\ref{EU-metric}) is estimated to reach the value:
\begin{equation}
\langle \Phi (r) \rangle = \frac 1V \int dV ~\Phi (r) = \frac {3\Omega}{d^5} \int_0^d dr r^4 \approx 0.6~.
\end{equation}
Thus the factor which would multiply spatial components of the metric in matter Lagrangians will be of the order of
\begin{equation} \label{S}
S^2 = \frac {1}{1 - \langle \Phi (r) \rangle} \approx 2.5~.
\end{equation}
Unlike the case of Friedmann-Robertson-Walker scale factor, one cannot hide the function $S$ in the definitions of the spatial coordinates, which are already set by the assumption to have the asymptotical Minkowski (or de Sitter) metric.

In the case of static, isotropic metrics (\ref{EU-metric}) one can apply the well-established effective potential technique to particle models. The optical-mechanical analogy leads to a remarkable simplification of the equations of motion of particles and general-relativistic problems become formally identical to the classical ones with the effective index of refraction \cite{refraction},
\begin{equation} \label{Refr}
n(r) = \sqrt{\frac {g_{rr}}{g_{tt}}} = \frac 1S \approx 0.6~.
\end{equation}
In addition to the gravitational refraction index (\ref{Refr}), in the Standard Model Lagrangian one can take into account the constant gravitational factor (\ref{S}) by introduction of the new time parameter, $t \to St$, and conducting the conformal transformation of the Minkowski metric,
\begin{equation} \label{Conformal}
\eta_{\mu\nu} \to \frac {1}{S^2} ~\eta_{\mu\nu}~,~~~~~ \sqrt {-\eta} \to \frac {1}{S^4} \sqrt {-\eta} ~.
\end{equation}
It is known that it is possible to bring the conformally equivalent matter Lagrangians to the Minkowskian form by the rescalings,
\begin{equation}
A^\nu \to A^\nu~, ~~~~~ \phi \to S \phi~, ~~~~~ \psi \to S^{3/2} \psi~,
\end{equation}
of the gauge, scalar and spinor fields, respectively \cite{GMM, Bir-Dav}.

In general the gravitational field affects strongly the symmetry behaviors of all quantum-field models, including scalar field lagrangians \cite{GMM, Bir-Dav}. Currently, we do not have a standard theory of massive scalar bosons in curved space-time, several models exist with the minimal or conformal couplings with curvature. Non-minimal couplings usually is supposed in inflation models \cite{Linde}. Investigations of the minimal case is important as well, since massive vector mesons and gravitons satisfy equations of this type \cite{GMM} and also for non-minimal couplings in general it is impossible to conserve conformal invariance not only of the effective action (the conformal anomaly) but of the action itself \cite{Bir-Dav}.

To explore properties of cosmological phase transitions in the presence of static external gravitational field, one should evaluate the expectation value of the Higgs field over the lowest energy state. To find an energy spectrum it is important construction of the Hamiltonian of the system, what in general is difficult problem in the presence of gravitational field. However, in stationary conformal metric there exist some eligible models, even for massive scalar fields \cite{Silenko}. It was found that in most situations, in regions where the conformal factor is almost constant, the conformal transformations finally amounts to a rescaling of a scalar boson mass in this region \cite{GMM, Bir-Dav, Silenko}. Our situation with the static gravitational field in some finite region of bulk space-time differs with the cases were the parameters of cosmological phase transitions were investigated in the infinite universe at the one-loop level \cite{Stable}.

Since a solution of the Higgs equation in the metric $\eta_{\mu\nu}$ is the solution of the similar equation with an effective mass (for simplicity we consider minimal coupling of the Higgs field to gravity) in the conformal metric, then under the transformation (\ref{Conformal}) the Standard Model Lagrangian  obtains the ordinary form, but with the modified Higgs mass,
\begin{equation} \label{m_H}
m_H \to \frac 1S~m_H \approx 0.6 ~ m_H ~.
\end{equation}
This means that in the early universe at the electroweak scale (in the symmetric phase) the effective vacuum expectation value (\ref{v}) probably was smaller than in the present broken phase,
\begin{equation} \label{v/S}
v \to 0.6 ~v \approx 156 ~ {\rm GeV} ~.
\end{equation}
Under the conformal rescalings (\ref{Conformal}) other parameters of the Standard Model are unchanged (including gauge boson masses), and in the perturbative analysis on the static background the modification of the vacuum expectation value (\ref{v/S}) leads to the alteration of the parameter (\ref{E}),
\begin{equation} \label{E/0.2}
E \to \frac {E}{0.2} \approx 0.05~.
\end{equation}
Then from the condition to have the first order electroweak phase transition in the early universe (\ref{l=2E}), we obtain the acceptable value for the Higgs self-coupling constant,
\begin{equation} \label{l}
\lambda \simeq 0.1~.
\end{equation}
This means that for the minimal Standard Model in the Einstein static universe background (\ref{EU-metric}), the electroweak phase transitions can be of the first order. After the expansion of new phase bubbles and the passage to the Friedman-Robertson-Walker expansion of spacetime in the broken phase, the gravitational potential in the universe (\ref{Phi}) will tend to zero and all parameters of the Standard Model will get the present values.


Note that, together with the allowance of electroweak phase transition to be of the first order, gravitational effects are able to solve another problem of the Standard Model baryogenesis -- the smallness of $CP$-violating parameters.

It is known that, in general, gravitation can induce $CP$-violation processes \cite{Hawking, ABCF, Cline, Kenyon, Char, Ahl-Kir, LMP, Amel-Cam}. Since antimatter can be interpreted as an ordinary matter propagating backward in time, for non-stationary spacetime (like an expanding new phase bubble) the time and thus $CP$-violation could be accrued in a $CPT$ preserving framework \cite{Bar-Gog}. To explore gravitational $CP$-violations for static backgrounds as well, one can use the fact that some quantum mechanical effects depend upon the gravitational potential themselves, not only to its gradient. Equations of quantum particles with a gravitational interaction terms contain inertial and gravitational mass separately. If within a model inertial and gravitational masses are not equal then in fermions wavefunctions there can appear the mass dependent gravitationally induced $CP$-violating phases \cite{Gasper, Hal-Leu}.

It is known that topological defects could violate the Week Equivalence Principle and exhibit nontrivial gravitational features \cite{VS95}. Since it is impossible to surround any topological object by a spatial boundary, one cannot define their gravitational mass, $M$, by the integral from the zero-zero component of energy-momentum tensor, $T_{00} $, but needs to use the Tolman formula,
\begin{equation}
M = \int \sqrt{-g}dV \left(T_0^0 - T_1^1 - T_2^2 - T_3^3 \right) ~.
\end{equation}
From this formula it follows that, in spite of having large tensions, cosmic strings do not produce any gravitational force on the surrounding matter locally, while global monopoles, global strings and planar domain walls exhibited repulsive nature \cite{VS95}.

In the bubble scenario spacetime is separated into two manifolds by a specific topological object -- a spherical domain wall \cite{BKT87}. Spherical domain walls with the outer Schwarzschild metric can be gravitationally repulsive as well (with the negative mass parameter), when the time coordinate changes its direction on the bubble surface \cite{BG94}, or if one assumes that the asymptotic metric is non-Minkowskian \cite{BG97}. Having written the interior/exterior metrics in a static form (Einstein's static universe (\ref{EU-metric}) inside and the Schwarzschild outside), we must allow for the time coordinate to be different in each region, it need not match across the bubble. On the other hand the radial coordinate measures the proper size of the spheres of a spherically symmetric spacetime and therefore has to vary continuously across the bubble. Note that the reversion of time direction is equivalent to the introduction of a new family of negative tension bubbles \cite{Mar-Wes, LLLNP}.

Due to the violation of the Weak Equivalence Principle, particles penetrating a topological object undergo strong jump in the gravitational potential,  $[\Phi]$, and pick up the addition phases in their wavefunctions \cite{Stod}. The gravitationally induced phases $\phi_{ij}$ can be approximated by \cite{Ahl-Bur},
\begin{equation}
\phi_{ij} \sim [\Phi] \Delta m^2_{ij}~,
\end{equation}
where $\Delta m_{ij}$ are the mass differences between different flavors. So the large jump of the gravitational potential at the new phase bubble walls, $[\Phi]$, could significantly increase $CP$-violating effects within the electroweak baryogenesis scenario.


In summary, it is taken for granted that the minimal Standard Model is unable to explain the baryon asymmetry of the universe, since electroweak phase transition in the early universe is appeared to be of second order (without non-equilibrium processes) and the $CP$-violation parameter from the Cabibbo-Kobayashi-Maskawa matrix is too small. However, if for the description of the metastable state at the electroweak scale we replace the Friedman-Robertson-Walker spacetime with the Einstein static universe model, the strong static gravitational potential leads to the effective reduction of the Higgs vacuum expectation value, which is found to be compatible with the first order electroweak phase transition conditions. We also argue that gravitational effects could increase the $CP$-violating parameters for the particles crossing the new phase bubble walls, which are appearing in electroweak baryogenesis scenarios.

\vskip 3mm

\noindent
{\bf Acknowledgments:} I would like to thank the University of Bonn for hospitality. This research partially is supported by Volkswagenstiftung (contract no. 86260).


\end{document}